\documentclass[preprint,aps]{revtex4}

\usepackage[dvipdfmx]{graphicx}
\def\vc#1{\mbox{\boldmath $#1$}}
\usepackage{wick}
\usepackage[dvipdfmx]{color}
\usepackage{amsmath}

\begin{document}

\title{A nuclear matter calculation with the tensor optimized Fermi sphere method using central interaction}






\author{Taiichi Yamada$^{1}$, Takayuki Myo$^{2,3}$, Hisashi Horiuchi$^{3}$, Kiyomi Ikeda$^{4}$, and Hiroshi Toki$^{3}$ }

\affiliation{$^{1}$College of Science and Engineering, Kanto Gakuin University, Yokohama 236-8501, Japan} 

\affiliation{$^{2}$General Education, Faculty of Engineering, Osaka Institute of Technology, Osaka, Osaka 535-8585, Japan}

\affiliation{$^{3}$Research Center for Nuclear Physics (RCNP), Osaka University, Ibaraki, Osaka 567-0047, Japan}

\affiliation{$^{4}$RIKEN Nishina Center, Wako, Saitama 351-0198, Japan}

\begin{abstract}%
The tensor optimized Fermi sphere (TOFS) method is applied first for the study of the property of nuclear matter using the Argonne V4' $NN$ potential. 
In the TOFS method, the correlated nuclear matter wave function is taken to be a power-series-type of the correlation function $F$, where $F$ can induce central, spin-isospin, tensor, etc.~correlations.
This expression has been ensured by a cluster expansion theory.
In the TOFS calculation, we take into account the contributions from all the many body terms arising from the product of the nuclear-matter Hamiltonian $\mathcal{H}$ and $F$. 
It is found that the density dependence of the energy per particle in nuclear matter is reasonably reproduced, in comparison with other methods such as the Brueckner-Hartree-Fock (BHF) approach. 
\\
\\
{PACS numbers: 21.65.-f
}\\
\end{abstract}


\maketitle

\section{Introduction}
\label{sec:Introduction}

It is one of the intriguing issues in nuclear physics to figure out the properties of nuclear and neutron matter on the ground of a bare interaction among nucleons.
At high density, their information is closely related to the structure of neutron-star interiors~\cite{glendenning,raffelt96}.
On the other hand, $\alpha$-particle (quartet) condensation due to the strong quartet correlations is predicted to occur at lower density region of symmetric nuclear matter~\cite{roepke98,beyer00,takemoto04,sogo09,sogo10-1,sogo10-2}.
This phenomena is related to $\alpha$-cluster states in finite system, such as the Hoyle $0^{+}_{2}$ state in $^{12}$C, observed in the excited energy region of light nuclei~\cite{wildermuth77,ikeda80,tohsaki00,yamada11,horiuchi12}.

In 1950s Brueckner et al.~presented their famous theories, ie.\ the Brueckner theory~\cite{brueckner55_1,brueckner55_2,brueckner58} and Brueckner-Bethe-Goldstone theory~\cite{bethe56,bethe57,goldstone57}, which treat nuclear matter from the bare interaction among nucleons.
Since then many-body theories have been developed by several groups.
The typical approaches in the non-relativistic framework are given as follows:~The Brueckner-Hartree-Fock (BHF) approach~\cite{brueckner58,mahayux,baldo91,schulze95}, the Brueckner-Bethe-Goldstone (BBG) approach up to the third order in hole-line expansion~\cite{day67,day83,song98,baldo00,baldo01,sartor06}, the self-consistent Green's function (SCGF) method~\cite{dickhoff04,frick05,soma06,rios09},  the auxiliary field diffusion Monte Carlo (AFDMC)~\cite{gandolfi07,gandolfi09}, the Green's function Monte Carlo (GFMC)~\cite{carlson03}, the Fermi hypernetted chain (FHNC) method~\cite{iwamoto57,fantoni72,fantoni74,fantoni78,pandharipande79,akmal98}, the coupled-cluster (CC) method~\cite{baardsen13,hagen14}, the quantum Monte Carlo lattice calculation~\cite{abe09}, and so on.
Among them Akmal, Pandharipande, and Ravenhall evaluated the ground-state energies for nuclear matter with the FHNC method using the modern nuclear Hamiltonian composed of the Argonne V18 (AV18) two-nucleon interaction~\cite{wiringa95} and the Urbana IX three-nucleon interaction~\cite{pudliner95}, taking into account the boost effect caused by the relativistic kinematics~\cite{akmal98}.
On the other hand, in the relativistic framework, Brockmann and Machleidt have proposed the Dirac-Brueckner-Hartree-Fock  (DBHF) method~\cite{brockmann90}, where the meson exchange interaction (Bonn potential) was used for the two nucleon interaction.

A comparative study using the Argonne two-nucleon potentials (AV4', AV6', AV8', and AV18)~\cite{wiringa95} for the many-body theories proposed so far is useful to clarify the quantitative differences among the theoretical calculations for the symmetric nuclear matter and neutron matter.
It is noted that AV4' is the central-force type, AV6' consists of only the central and tensor forces, AV8' does of the central, tensor, and spin-orbit forces, and AV18 is one of the modern two-nucleon interaction expressed as a sum of 18 operators.
Such comparative study has been performed by Baldo et al.~\cite{baldo12} for BHF, BBG, SCGF, AFDMC, GFMC, and FHNC.
According to their paper, the difference of the density dependence of the energy per particle, $E(\rho)/A$, for the neutron matter among various methods is likely to be small for any Argonne potential, although the discrepancies among them are gradually enhanced at higher density.
In the symmetric nuclear matter, however, we see non-negligible dependence of the behavior of $E(\rho)/A$ on various methods.
For example, in the cases of AV8' and AV6', the significant dependence of the behavior of $E(\rho)/A$ on the calculated methods (BHF, BBG, SCGF, AFDMC, and FHNC) exist even in the lower density region including the normal density $\rho_0$, and they are amplified in higher density, although the dependence in the case of the AV4' potential (central force) is relatively small. 
The contribution from the non-central forces, in particular, tensor force (and spin-orbit one), is more important in the symmetric nuclear matter than the neutron matter.
Thus the dependence of $E(\rho)/A$ on the calculated methods is likely to be caused mainly by the different treatment of the non-central components in medium, in particular, tensor component (and spin-orbit one) in each methods, together with the treatment of many-body terms appearing in the nuclear matter calculations. 
The above mentioned facts suggest that there exists the matter of convergence of $E(\rho)/A$.  

Recently, the tensor-optimized antisymmetrized molecular dynamics (TOAMD) has been proposed~\cite{myo15,myo17_1,myo17_2,myo17_3,myo17_4,myo17_5}. 
This is a variational framework for ab initio description of light nuclei, where the AMD wave function~\cite{enyo03,enyo12} is used as the uncorrelated wave function, and the correlation functions for the central-operator and tensor-operator types, $F_S$ and $F_D$, respectively, are introduced and employed in power series form of the wave function.
In the analysis of {\it s}-shell nuclei with TOAMD, they have nicely reproduced the results of Green's function Monte Carlo (GFMC) by using the double products of the correlation functions, where each correlation function in every term is independently optimized in the variation of total energy.

It is interesting to seek a framework for describing nuclear matter with the power series correlated wave function.
In the FHNC framework, the Jastrow-type correlated wave function is used to describe the nuclear matter, and the energy per particle is evaluated in a diagrammatic cluster expansion.
This is based on the linked-cluster expansion theorem for the Jastrow-type correlated wave function proved by Fantoni and Rosani~\cite{fantoni72,fantoni74,fantoni78,pandharipande79}:~Only the linked diagrams contributes to the energy per particle in the nuclear matter and the unlinked ones are cancel out at each order of the cluster expansion.
In the TOAMD framework describing finite nuclei, one can take an arbitrary power series form with respect to the correlation functions, $F_S$ and $F_D$.
However, it is non-trivial whether any power series form is allowed or not for describing the nuclear matter from the light of the cluster expansion for the energy. 
One needs to explore a formalism satisfying a sort of the linked-cluster expansion theorem in the case of using the power series correlated wave function.

Quite recently, a new formalism, called ``tensor optimized Fermi sphere (TOFS ) method'', has been presented to treat the nuclear matter using a bare interaction among nucleons, together with a cluster expansion theory for TOFS~\cite{yamada18_cluster}.
In this formalism based on Hermitian form, the correlated nuclear matter wave function is taken to be a power series one, $\Psi_{N}=[\sum_{n=0}^{N} {(1/n!)F^n}]\Phi_0$ and an exponential one, $\Psi_{\rm ex}=\exp(F) \Phi_0$, where the correlation function $F$  with $F=F_S+F_D$ can induce the central correlation $F_S$ and the tensor correlation $F_D$, and $\Phi_0$ is the uncorrelated Fermi-gas wave function. 
$\Psi_{\rm ex}$ corresponds to a limiting case of $\Psi_{N}$ ($N \rightarrow \infty$).
It is noted that both the central and tensor correlations play an important role in nuclear matter as well as finite nuclei.
The cluster expansion theory for the TOFS formalism~\cite{yamada18_cluster} tells us that the energy per particle in nuclear matter with $\Psi_{\rm ex}$ is expressed in terms of a linked-cluster expansion.
Based on this fact, the formula of the energy per particle in nuclear matter with $\Psi_{N}$ is given as an approximation of that with $\Psi_{ex}$.
The correlation functions are optimally determined in variation of the energy of nuclear matter. 
The method of evaluating the energy of nuclear matter with $\Psi_N$ is called the $N$th order TOFS calculation.

In this paper, the 1st order TOFS method ($N=1$) is applied for the study of the properties of nuclear matter with using the Argonne V4' two-body potential (central-force-type).
In this calculation, we take into account all of the contributions from the one-body to six-body terms arising from the correlated Hamiltonian composed of the correlation function $F$ and the nuclear matter Hamiltonian $\mathcal{H}$.
The correlation functions are expanded into the Gaussian functions with appropriate Gaussian ranges, and their expansion coefficients are optimally determined in the variation of the energy.
The calculated results are reasonably reproduced compared with those of BHF, BBG, SCGF, AFDMC etc.~, as shown later.
This indicates that the present TOFS method deserves to evaluate the binding energy per particle in nuclear matter and neutron matter with the more complex bare interaction among nucleons.
    
The present paper is organized as follows:~In Sec.~\ref{sub:brief_TOFS}, we give the formulation of the TOFS method, emphasizing the relationship between the TOFS method and cluster expansion theorem.
The 1st order TOFS method is formulated in Sec.~\ref {sub:1st-order_TOFS}.
We demonstrate that the binding energy per particle in nuclear matter and neutron matter is obtained by solving a linear equation of the Gaussian expansion coefficients of the correlation functions.\
The calculated results and discussion are presented in Sec.~\ref{sub:results_and_discussion}.
Finally, we give a summary in Sec.~\ref{sub:summary}.

\section{Formulation of TOFS method}
\label{sub:brief_TOFS}

In this section we present a brief formulation of the TOFS method for nuclear matter with a bare interaction among nucleons.
The details of the TOFS method and cluster expansion theory are given in Ref.~\cite{yamada18_cluster}.

In the TOFS framework, we consider the two types of the correlated wave function (wf) of symmetric nuclear matter.
One is the $N$th-order power-series-type TOFS wf, $\Psi_N$,  and the other is the exponential-type TOFS wf, $\Psi_{\rm ex}$, which is the limiting case of $N \to \infty$ for $\Psi_N$,
\begin{eqnarray}
&&\Psi_N = \left[ \sum_{k=0}^{N} \frac{1}{k!} F^{k} \right] \Phi_0, 
\label{eq:correlated_wf_finite}
\\
&&\Psi_{\rm ex} = \exp(F) \Phi_0 = \lim_{N \to \infty }\,\Psi_N,
\label{eq:correlated_wf_exp}
\\
&&F = F_S + F_D.
\label{eq:correlation_fun}
\end{eqnarray}
The uncorrelated wave function $\Phi_0$ in Eqs.~(\ref{eq:correlated_wf_finite}) and (\ref{eq:correlated_wf_exp}) is described by the Fermi  gas model, in which $A$ nucleons occupy up to the Fermi sea with the Fermi wave number $k_F$, 
\begin{eqnarray}
\Phi_{0} = \frac{1}{\sqrt{A!}}\ {\det}\ | \phi_{\gamma_1}(1) \phi_{\gamma_2}(2) \cdots \phi_{\gamma_A}(A)|.
\label{eq:femi_gas_wf}
\end{eqnarray}
The single-nucleon wave function $\phi_\gamma$ confined in a box with length $L$ and volume $\Omega=L^3$ is written as
\begin{eqnarray}
\begin{split}
&\phi_{\gamma_n}(n) = \phi_{\vc{k}_n}(\vc{r}_n)\ \chi_{m_{s_n}}(n)\  \xi_{m_{t_n}}(n),\\
&\phi_{\vc{k}_n}(\vc{r}_n) = \frac{1}{\sqrt{\Omega}} \exp ( i \vc{k}_n \cdot \vc{r}_n), 
\end{split}
\label{eq:spwf}
\end{eqnarray}
where $\gamma=(\vc{k},m_s,m_t)$ represents the quantum number of the single-nucleon wave function, and $\chi$ and $\xi$ are the spin and isospin wave functions, respectively.
The single-nucleon wave function in Eq.~(\ref{eq:spwf}) is imposed the periodic boundary condition. 
Then, we obtain $\rho = 2k_{F}^{3}/(3\pi^2)$, where we consider the infinite nuclear matter in the limit of large $L$ (or $\Omega$).

The correlation functions $F_S$ and $F_D$ in Eq.~(\ref{eq:correlation_fun}), which describe the spin-isospin dependent centrals correlation and tensor correlation in nuclear matter, respectively, are defined as
\begin{eqnarray}
F_S &=& \frac{1}{2} \sum_{i  \not= j} f_S(i,j) = \frac{1}{2} \sum_{s=0}^{1}\sum_{t=0}^{1}\sum_{i \not= j} f_S^{(st)}(r_{ij})P^{(st)}_{ij}, 
\label{eq:fs}\\
F_D &=& \frac{1}{2} \sum_{i \not=j} f_D(i,j) = \frac{1}{2} \sum_{s=0}^{1}\sum_{t=0}^{1} \sum_{i \not= j}f_D^{(st)} (r_{ij}) r^2_{ij} S_{12}(i,j)P^{(st)}_{ij} \delta_{s1},
\label{eq:fd} 
\end{eqnarray}
where $S_{12}$ is the tensor operator, 
\begin{eqnarray}
&&S_{12}(i,j)= 3(\vc{\sigma}_i\cdot\hat{\vc{r}}_{ij})(\vc{\sigma}_j\cdot\hat{\vc{r}}_{ij}) - (\vc{\sigma}_i\cdot\vc{\sigma}_j)
\end{eqnarray}
with $\hat{\vc{r}}_{ij}=\vc{r}_{ij}/r_{ij}$ and $\vc{r}_{ij}=\vc{r}_{i} - \vc{r}_{j}$.
The operator $P^{(st)}_{ij}$ denotes the projection operator of the spin $s$ and isospin $t$ states of the $ij$-nucleon pair:~$P^{(st)}_{ij}=P^{(s)}_{ij} P^{(t)}_{ij}$ with 
\begin{eqnarray}
&&P^{(s)}_{ij}=\frac{1}{4}\left[ (2s+1)+(-1)^{s+1}(\vc{\sigma}_i \cdot \vc{\sigma}_j)\right], 
\label{eq:Ps}\\
&&P^{(t)}_{ij}=\frac{1}{4}\left[ (2t+1)+(-1)^{t+1}(\vc{\tau}_i \cdot \vc{\tau}_j)\right],
\label{eq:Pt}
\end{eqnarray}
where the spin operators $\vc{\sigma}_i$ and $\vc{\sigma}_j$ (isospin operators $\vc{\tau}_i$ and $\vc{\tau}_j$) are for the particles {\it i} and {\it j}, respectively.
One may add other type correlation functions such as a spin-orbit-type correlation function $F_{SO}$ into $F$ in Eq.~(\ref{eq:correlation_fun}) as needed, $F=F_{S}+F_{D}+F_{SO}$.

The Hamiltonian of the nuclear matter is expressed as a sum of the kinetic energies, two-body interactions, and three-body interactions,
\begin{eqnarray}
\mathcal{H} = -\frac{\hbar^2}{2m} \sum_{i=1}^{A} \vc{\nabla}_i^2 + \frac{1}{2} \sum_{i  \not= j}^{A} v(i,j) + \frac{1}{6} \sum_{i \not= j \not= k}^{A} V(i,j,k),
\label{eq:hamiltonian}
\end{eqnarray}
where $m$ denotes the nucleon mass.
According to the cluster expansion theory under the present TOFS method discussed in Ref.~\cite{yamada18_cluster}, the binding energy per particle in nuclear matter, $B_{\rm ex}$, with use of the exponential-type correlated wf $\Psi_{\rm ex}$ in Eq.~(\ref{eq:correlated_wf_exp}) is presented as 
\begin{eqnarray}
&&-B_{\rm ex} = \frac{1}{A}\, \frac{\langle \Psi_{\rm ex} | \mathcal{H} | \Psi_{\rm ex} \rangle }{\langle \Psi_{\rm ex} | \Psi_{\rm ex}  \rangle }  =  \frac{1}{A}\, \frac{\left\langle \Phi_0 |  \exp(F^{\dagger}) \mathcal{H} \exp(F)  | \Phi_0 \right\rangle}{\left\langle \Phi_0 | \exp(F^{\dagger}) \exp(F)  | \Phi_0 \right\rangle}
= \frac{1}{A}\, \sum_{n=0}^{\infty}\, (E_n)_{\rm c},
\label{eq:B_ex}
\\
&&(E_n)_{\rm c} = \sum_{\substack{n_1,n_2\\{n_1+n_2=n}}} \frac{1}{{n_1}!\  {n_2}!}\ {\left\langle \Phi_0 | F^{n_1}\mathcal{H}F^{n_2} | \Phi_0 \right\rangle}_{\rm c},
\label{eq:Enc}
\end{eqnarray}
where ${\left\langle \Phi_0 | F^{n_1}\mathcal{H}F^{n_2} | \Phi_0 \right\rangle}_{\rm c}$ is the summation of the linked diagrams in the matrix element of ${\left\langle \Phi_0 | F^{n_1}\mathcal{H}F^{n_2} | \Phi_0 \right\rangle}$ (see Ref.~\cite{yamada18_cluster} and also Sec.~\ref{subsec:formulation} in this paper), and 
$(E_n)_{\rm c}$ stands for the $n$th-order cluster energy.
It is noted that the unlinked diagrams in each matrix element are completely canceled out in each order in the cluster expansion, and then only linked diagrams remains as shown in Eq.~(\ref{eq:Enc}). 
The expression (\ref{eq:B_ex}) means that the binding energy per particle, $B_{\rm ex}$,  is expressed as the sum of only the linked diagrams.
Since each integral of the linked diagram is proportional to the number $A$, $B_{\rm ex}$ becomes $A$-independent or $\rho$-dependent~\cite{yamada18_cluster}.
All the linked diagrams~\cite{yamada18_cluster} can be classified into the two classes, `\textit{simple}' and  `\textit{composite}', and the former can be decomposed into two groups, `\textit{nodals}' and `\textit{elementary}', as in the case of the FHNC framework~\cite{fantoni72,fantoni74,fantoni78,pandharipande79}.  

In the case of taking the $N$th-order power series wave function $\Psi_{N}$ in Eq.~(\ref{eq:correlated_wf_finite}) as the correlated wave function of nuclear matter, the binding energy per nucleon, $B_{N}$, is expressed as an approximation of $B_{\rm ex}$ in Eq.~(\ref{eq:B_ex}),
\begin{eqnarray}
&&{-B}_{N}
= \frac{1}{A}\,\frac{\left\langle \Psi_{N} \left| \mathcal{H} \right|  \Psi_{N} \right\rangle}{\left\langle \Psi_{N} |  \Psi_{N} \right\rangle} 
\simeq
\frac{1}{A}\,\sum_{n_1=0}^{N} \sum_{n_2=0}^{N} \frac{1}{{n_1}!\  {n_2}!}\ {\left\langle \Phi_{0} \left| F^{n_1}\mathcal{H}F^{n_2} \right| \Phi_{0} \right\rangle}_{\rm c},
\label{eq:BE_linked}
\end{eqnarray}
where only the linked diagrams in the matrix element are evaluated~\cite{yamada18_cluster}.
We notice that the right side in Eq.~(\ref{eq:BE_linked}) is independent of $A$ or depends on only $\rho$, and converges definitely to $-B_{\rm ex}$ in Eq.~(\ref{eq:B_ex}) in the limit of $N \rightarrow \infty$.
Since the Hamiltonian $\mathcal{H}$ in Eq.~(\ref{eq:hamiltonian}) has one-body, two-body, and three-body operators, the product operator $F^{n_1}\mathcal{H}F^{n_2}$ is expressed as the summation from the one-body to $(2n_1+2n_2+3)$-body operators. 
The explicit expressions of $B_N$ with $N=1,2$ are given as
\begin{eqnarray}
\begin{split}
-B_{N=1} &= \frac{1}{A}\,\left[{\langle \Phi_0 | \mathcal{H} | \Phi_0 \rangle}_{\rm c} +  {\langle \Phi_0 | F\mathcal{H} + \mathcal{H}F | \Phi_0 \rangle}_{\rm c} +  {\langle \Phi_0 | F\mathcal{H}F | \Phi_0 \rangle}_{\rm c}\,\right], 
\end{split}
\label{eq:1st_TOFS_cal}
\end{eqnarray}
\begin{eqnarray}
&&-B_{N=2} = \frac{1}{A} \left[ {\langle \Phi_0 | \mathcal{H} | \Phi_0 \rangle}_{\rm c} +  {\langle \Phi_0 | F\mathcal{H} + \mathcal{H}F | \Phi_0 \rangle}_{\rm c} +  {\left\langle \Phi_0 \left| \frac{1}{2!}F^2\mathcal{H} + F\mathcal{H}F + \frac{1}{2!} \mathcal{H}F^2 \right| \Phi_0 \right\rangle}_{\rm c} \right. \nonumber \\
&&\hspace*{25mm}\left.+ {\left\langle \Phi_0 \left| \frac{1}{2!}F^2\mathcal{H}F + \frac{1}{2!} F\mathcal{H}F^2 \right| \Phi_0 \right\rangle}_{\rm c}
+ {\left\langle \Phi_0 \left| \frac{1}{{2!}^2}F^2\mathcal{H}F^2 \right| \Phi_0 \right\rangle}_{\rm c}\,\right],
\label{eq:2nd_TOFS_cal}
\end{eqnarray}
where $F=F_S + F_D$ given in Eq.~(\ref{eq:correlation_fun}).
We call the 1st order (2nd order) TOFS calculation for evaluating $B_{N=1}$ ($B_{N=2}$). 

In the present TOFS framework, the radial parts of the correlation functions in Eqs.~(\ref{eq:fs}) and ($\ref{eq:fd}$) are expanded in terms of the Gaussian functions,
\begin{eqnarray}
&&f_S^{(st)}(r) = \sum_{\mu}C^{(st)}_{S,\mu} \exp\left[-a^{(st)}_{S,\mu} r^2\right],
\label{eq:exp_gs_FS}\\
&&f_D^{(st)}(r) = \sum_{\mu}C^{(st)}_{D,\mu} \exp\left[-a^{(st)}_{D,\mu} r^2\right].
\end{eqnarray}
Here, $C^{(st)}_{S,\mu}$ and $a^{(st)}_{S,\mu}$ together with $C^{(st)}_{D,\mu}$ and $a^{(st)}_{D,\mu}$ are the variational parameters.
They are determined so as to minimize the energy per particle in nuclear matter.
It is noted that the Gaussian correlation functions bring about simplification and numerical stabilization for evaluating the matrix elements of many-body operators in the present nuclear matter calculation.  
This Gaussian expansion method of the correlated functions is the same as that done in finite nuclear calculations with TOAMD~\cite{myo15,myo17_1,myo17_2,myo17_3,myo17_4,myo17_5,myo17_6,lyu17}.    
The values of the size parameters, $a^{(st)}_{S,\mu}$  and $a^{(st)}_{D,\mu}$, are appropriately chosen by considering the range of the nuclear force and the Fermi wave number $k_F$, and only the expansion coefficients, $\left\{C_{S,\mu}^{(st)}\right\}$ and $\left\{C_{D,\mu}^{(st)}\right\}$, are taken as the variational parameters.
They are determined from the following conditions:
\begin{eqnarray}
\frac{\partial B_{N}}{\partial C^{(st)}_{S,\mu}}=0,\ \ \ \ \ 
\frac{\partial B_{N}}{\partial C^{(st)}_{D,\mu}}=0. 
\label{eq:system_eq}
\end{eqnarray}
The solutions to the system of equations $(\ref{eq:system_eq})$ for $C^{(st)}_{S,\mu}$ and $C^{(st)}_{D,\mu}$ give the variational minimization of the energy per particle in nuclear matter, $-B_N$, in Eq.~(\ref{eq:BE_linked}).

In the $N$-th order TOFS calculation, the effects of many-body correlations originating from the correlated Hamiltonian $\left({\sum_{n=0}^N \frac{1}{n!} {F}^n}\right) \mathcal{H} \left({\sum_{n=0}^N \frac{1}{n!} F^n}\right)$ in Eq.~(\ref{eq:BE_linked}) are taken into account as the follows:~All the contribution from one-body to $(2N+3)$-body terms arising from the product operator $\left({\sum_{n=0}^N \frac{1}{n!} {F}^n}\right) \mathcal{H} \left({\sum_{n=0}^N \frac{1}{n!} F^n}\right)$ are evaluated in the TOFS calculation.
As a result one can include all the `\textit{elementary}' and `\textit{nodals}' diagrams together with the `\textit{composite}' ones which appear in calculating the matrix element in Eq.~(\ref{eq:BE_linked}).   
This is different from the variational method based on the hypernetted chain summation techniques (VCS or FHNC-SOC) by Pandharipande et al.~\cite{pandharipande79,akmal98}.
In the VCS, the many-body correlations are taken into account through the diagrams described by the single operator chain (SOC) approximation in the `\textit{nodals}' diagrams by the SOC equations, although the elementary diagrams are neglected.
The computations using the correlated basis function theory and Fermi hypemetted chain theory have been performed for the finite nuclei ~\cite{co94,arias07}.
Their results indicate that the effect of the elementary diagrams is not negligible in the binding energies of the ground states of $^{16}$O and $^{40}$Ca.
Thus, the quantitative estimation for the effect of the elementary diagrams seems to be of more importance in nuclear matter, presumably of importance in dense nuclear matter. 

\section{First order TOFS calculation with the central force}
\label{sub:1st-order_TOFS}

In this section we formulate the 1st order TOFS method of calculation with the central force.

\subsection{Formulation}
\label{subsec:formulation}

The $A$-body Hamiltonian with the two-body $NN$ central force is written as 
\begin{eqnarray}
\begin{split}
\mathcal{H} 
&= T + V_{C} \\
&= \sum_{i=1}^{A}\,t(i)+ \frac{1}{2} \sum_{i \not=j}^{A} \left[ \sum_{s=0}^{1} \sum_{t=0}^{1} v_{C}^{(st)}(r_{ij}) P^{(st)}_{ij} \right],
\end{split}
\label{eq:hamiltonian_central}
\end{eqnarray}
where the spin-isospin projection operator $P^{(st)}_{ij}=P^{(s)}_{ij} P^{(t)}_{ij}$ is defined in the previous section.
For the correlated nuclear matter wave function, we take the 1st order TOFS wave function
\begin{eqnarray}
\Psi_{N=1} = \left[ 1 + F_{S} \right] \Phi_{0},
\label{eq:wf_1st_TOFS}
\end{eqnarray}
where $F_{S}$ stands for the spin-isospin dependent central correlation function, defined in Eq.~(\ref{eq:fs}).
It is noted that the tensor-force type correlated function $F_D$ is omitted, because only the central force is relevant in the present study.  
According to Sec.~\ref{sub:brief_TOFS} and Ref.~\cite{yamada18_cluster}, the binding energy per particle in nuclear matter $B_{N=1}$ in the 1st order TOFS calculation is given in Eq.~(\ref{eq:1st_TOFS_cal}). 

In order to evaluate $B_{N=1}$, we have to calculate the matrix element $\langle \Phi_{0} | {F_S}^{n_1} \mathcal{O} {F_S}^{n_2} | \Phi_{0} \rangle$ with $0 \le n_1+n_2 \le 1$, and $\mathcal{O}=T$ and $V_{C}$.
This matrix element can be divided into the following two terms,
\begin{eqnarray}
{\langle \Phi_{0} | {F_S}^{n_1} \hat{O} {F_S}^{n_2} | \Phi_{0} \rangle} = {\langle \Phi_{0} | {F_S}^{n_1} \hat{O} {F_S}^{n_2} | \Phi_{0} \rangle}_{\rm c} + {\langle \Phi_{0} | {F_S}^{n_1} \hat{O} {F_S}^{n_2} | \Phi_{0} \rangle}_{\rm dis}.   
\label{eq:c_dis_FOF}
\end{eqnarray}
The first term in the right side, called {\it linked matrix element}, denotes the sum of all the linked diagrams (integrals) in the matrix element ${\langle \Phi_{0} | {F_S}^{n_1} \hat{O} {F_S}^{n_2} | \Phi_{0} \rangle}$.
The second term, called {\it unlinked matrix element}, is that of all the unlinked diagrams (integrals) in the matrix element.
Although the definition of the linked (unlinked) matrix element is discussed in detail in Ref.~\cite{yamada18_cluster}, we will briefly explain them below.  

The product operator $F^{n_1} \mathcal{O} F^{n_2}$ in Eq.~(\ref{eq:c_dis_FOF}) can be decomposed into the sum of the multi-body operators.
For example,  $F_S\,T$ ($F_S\,V_{C}$, $F_S\,T\, F_S$, and $F_S\,V_{C}\, F_S$) has the multi-body operators from two-body to three-body one (four-body, five-body, and six-body ones, respectively)~\cite{myo15},  
\begin{eqnarray}
\begin{split}
F_{S}\, T &= \left( \frac{1}{2} \sum_{i \not= j} f_{S}(i,j) \right) \left( \sum_{i} t(i) \right), \\
&= {\sum_{i \not= j} f_{S}(i,j) t(i)} + \frac{1}{2} {\sum_{i \not=j \not= k} f_{S}(i,j) t(k) } \equiv 1(12)(1) + \frac{1}{2}(12)(3),
\end{split}
\label{eq:FT}
\end{eqnarray}
\begin{eqnarray}
\begin{split}
F_{S}\, V_{C} &= \left( \frac{1}{2} \sum_{i \not= j} f_{S}(i,j) \right) \left( \frac{1}{2} \sum_{i \not= j} v(i,j) \right), \\
&= \frac{1}{2} {\sum_{i \not= j} f_{S}(i,j) v(i,j)} + {\sum_{i \not=j \not= k} f_{S}(i,j) v(i,k) }  + \frac{1}{4} {\sum_{i \not=j \not= k \not= l} f_{S}(i,j) v(k,l) }, \\
&\equiv \frac{1}{2} (12)^2 + 1(12)(13) + \frac{1}{4}(12)(34),
\end{split}
\label{eq:FV}
\end{eqnarray}
\begin{eqnarray}
\begin{split}
F_{S}\, T\,F_{S} &=  \left( \frac{1}{2} \sum_{i \not= j} f_{S}(i,j) \right)   \left( \sum_{i} t(i) \right) \left( \frac{1}{2} \sum_{i \not= j} f_S(i,j) \right), \\
&\equiv 1(12)(1)(12) + 1(12)(1)(13) + 1(12)(1)(23) + \frac{1}{2}(12)(3)(12) + 1(12)(3)(23) \\
&\hspace*{2mm} + \frac{1}{2}(12)(1)(34) + \frac{1}{2}(12)(3)(34) + 1(12)(3)(14) \\
&\hspace*{2mm} +\frac{1}{4}(12)(3)(45),
\end{split}
\label{eq:FTF}
\end{eqnarray}
\begin{eqnarray}
\begin{split}
F_{S}\, V_{C}\,F_{S} &=  \left( \frac{1}{2} \sum_{i \not= j} f_{S}(i,j) \right) \left( \frac{1}{2} \sum_{i \not= j} v(i,j) \right)  \left( \frac{1}{2} \sum_{i \not= j} f_{S}(i,j) \right), \\
&\equiv \frac{1}{2} (12)^3 + 1(12)^2(13) + 1(12)(13)(12) +1(12)(13)^2 + 1(12)(13)(23) \\
&\hspace*{2mm} + \frac{1}{4}(12)^2(34) + 1(12)(13)(14) + 1(12)(13)(24) +1(12)(13)(34) \\
&\hspace*{10mm} + \frac{1}{4}(12)(34)(12) + 1(12)(34)(13) + \frac{1}{4}(12)(34)^2 \\
&\hspace*{2mm} + \frac{1}{2}(12)(13)(45) + \frac{1}{2}(12)(34)(15) + \frac{1}{2}(12)(34)(35) \\
&\hspace*{2mm} + \frac{1}{8}(12)(34)(56).
\end{split}
\label{eq:FVF}
\end{eqnarray}

\begin{figure}[t]
\centering
\includegraphics*[width=0.6\hsize]{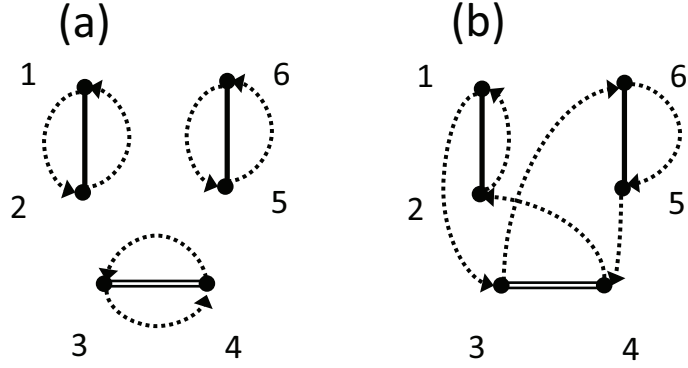}
\caption{
Examples of the (a) unlinked and (b) linked diagrams in the matrix element of the 6-body operator, $\tfrac{1}{8}(12)(34)(56)$, of the product operator $F_S V_C F_S$. The solid line and double line denote the correlation function $f_S$ and nuclear force $v_C$, respectively, and the dotted lines represent $g_\beta$. See the text.
}
\label{fig:diagrams_unlinked_linked}
\end{figure} 

Here it is instructive to discuss the linked and unliked diagrams appeared in the matrix element of the six-body operator in $F_S V_C F_S$ in Eq.~(\ref{eq:FVF}), as an example, shown as 
\begin{eqnarray}
\frac{1}{8}\,(12)(34)(56) = \frac{1}{8}\,\sum_{i\not=j\not=k\not=l\not=m\not=n}f_{S}(i,j)v(k,l)f_{S}(m,n).
\label{eq:6-body}
\end{eqnarray}
The explicit expression of its matrix element~\cite{yamada18_cluster} is given as
\begin{eqnarray}
{\left\langle\, \Phi_{0}\, \left|\, \frac{1}{8}(12)(34)(56)\, \right|\, \Phi_{0}\, \right\rangle} 
= A\times \rho^5 \times \sum_{\beta}\,{\rm sgn}(\beta)\,\int d\{\vc{r}\} \,G_{\beta}(\{\vc{r}\}), 
\label{eq:mt_123456}\\
\begin{split}
G_{\beta}(\{\vc{r}\}) &= \frac{1}{8}\ \sum_{s_1,t_1} \sum_{s_2,t_2} \sum_{s_3,t_3}\,
\frac{1}{4^6}\, F_{\beta}^{(6)(12:34:56)}(s_1,s_2,s_3) \, F_{\beta}^{(6)(12:34:56)}(t_1,t_2,t_3) \\
&\times f_{S}^{(s_1t_1)}(r_{12})\, v_{C}^{(s_2t_2)}(r_{34})\, f_{S}^{(s_3t_3)}(r_{56})\,g_{\beta}^{(6b)}(\{ \vc{r} \}),
\end{split}
\\
\begin{split}
g_{\beta}^{(6b)}(\{ \vc{r} \}) = \prod_{n=1}^{6} h(k_F\rho_n(\beta)),\ \ \ \ h(x)=\frac{3j_1(x)}{x},\hspace*{35mm}
\end{split}
\label{eq:h(x)}
\\
\beta=
\begin{pmatrix}
1\ \ 2 & 3\ \ 4 & 5\ \ 6 \\
\beta_1\ \beta_2 & \beta_3\ \beta_4 & \beta_5\ \beta_6 \\
\end{pmatrix},
\hspace*{50mm}
\end{eqnarray}
where $\beta$ represents the permutation for the numbers from 1 to 6, having totally $6!=720$, ${\rm sgn}(\beta)$ denoting the signature of $\beta$, $\vc{\rho}_{n}(\beta)=\sum_{j=1}^{6}\delta_{n\beta_j}\vc{r}_j -\vc{r}_n$, $(\{\vc{r}\})=(\vc{r}_{12},\vc{r}_{34},\vc{r}_{56},\vc{r}_{15},\vc{r}_{45})$, and $d\{\vc{r}\}=d\vc{r}_{12}d\vc{r}_{34} d\vc{r}_{56} d\vc{r}_{15} d\vc{r}_{45}$.
The spin-isospin matrix element $F_\beta$ is defined as
\begin{eqnarray}
\begin{split}
F_\beta^{(6)(12:34:56)}(s_1,s_2,s_3) 
= \sum_{\substack{m_1, m_2, m_3,\\ m_4, m_5, m_6}} \left\langle m_1 m_2 m_3 m_4 m_5 m_6 \left| P_{12}^{(s_1)} P_{34}^{(s_2)} P_{56}^{(s_3)} \right| {m\{\beta\}} \right\rangle,
\end{split}
\end{eqnarray}
where $m_1=\pm\tfrac{1}{2}$, $m_2=\pm\tfrac{1}{2}$, $\cdots$, and $m_6=\pm\tfrac{1}{2}$, and $| m\{\beta\} \rangle = |m_{\beta_1}m_{\beta_2}\cdots m_{\beta_6} \rangle$.
It is noted that $g_\beta$ and $F_\beta$ are related to the antisymmetrization among nucleons.

The structure of each integrand $G_\beta$ in Eq.~(\ref{eq:mt_123456}) can be represented diagrammatically~\cite{yamada18_cluster}.
They are classified into linked and unlinked diagrams, depending on the permutation $\beta$ and the type of the multi-body operator. 
For instance, let's consider the permutation
$\beta=
\begin{pmatrix}
1\ 2 & 3\ 4 & 5\ 6 \\
2\ 1 & 4\ 3 & 6\ 5 \\
\end{pmatrix}
$
for the multi-body operator in Eq.~(\ref{eq:6-body}).
This diagrammatic representation is shown in Fig.~\ref{fig:diagrams_unlinked_linked}(a).
In this case the diagram is unlinked, because the permutation is presented by the product of the three sub-permutations,
$\begin{pmatrix}
1\ 2 & 3\ 4 & 5\ 6 \\
2\ 1 & 4\ 3 & 6\ 5 \\
\end{pmatrix}
=
\begin{pmatrix}
1\ 2 \\
2\ 1 \\
\end{pmatrix}
\begin{pmatrix}
3\ 4 \\
4\ 3 \\
\end{pmatrix}
\begin{pmatrix}
5\ 6 \\
6\ 5 \\
\end{pmatrix}
$.
Then the integral $\int d\{\vc{r}\} \,G_{\beta}(\{\vc{r}\})$, which is proportional to $\Omega^2$, is divergent in infinite system, where $\Omega$ denotes the volume of nuclear matter.
On the other hand, in the case of $\beta=
\begin{pmatrix}
1\ 2 & 3\ 4 & 5\ 6 \\
3\ 1 & 6\ 2 & 4\ 5 \\
\end{pmatrix}
$,
the diagrammatic representation of $G_\beta$ is linked, as shown in Fig.~\ref{fig:diagrams_unlinked_linked}(b).
Then, the integral $\int d\{\vc{r}\} \,G_{\beta}(\{\vc{r}\})$ for this linked diagram is not divergent.

The 720 integrals in Eq.~(\ref{eq:mt_123456}) can be easily classified into the two-type diagrams, linked and unlinked.
As a result, we have the 592 linked integrals and 128 unlinked ones. 
The total sum of the linked (unlinked) diagrams appearing in Eq.~(\ref{eq:mt_123456}) is called the linked (unlinked) matrix element for the 6-body operator in $F_{S} V_{C} F_{S}$.  
This classification is applicable for other types of the multi-body operator in $F_S V_C F_S$ as well as those for $F^{n_1}\mathcal{O}F^{n_2}$ with $0 \le n_1, n_2 \le 1$ and $\mathcal{O}=T, V_C$.
Consequently one can divide the matrix element $\langle \Phi_0 | {F_S}^{n_1}\mathcal{O}{F_S}^{n_2} | \Phi_0 \rangle$ into the linked and unlinked matrix elements, $\langle \Phi_0 | {F_S}^{n_1}\mathcal{O}{F_S}^{n_2} | \Phi_0 \rangle_{\rm c}$ and $\langle \Phi_0 | {F_S}^{n_1}\mathcal{O}{F_S}^{n_2} | \Phi_0 \rangle_{\rm dis}$, respectively, as shown in Eq.~(\ref{eq:c_dis_FOF}).
 
The radial part of the correlation function $F_S$ is expanded into the Gaussian function, as shown in Eq.~(\ref{eq:exp_gs_FS}).
Then the correlation function $F_S$ can be written down as
\begin{eqnarray}
\begin{split}
&F_{S} = \sum_{s,t,\mu} C_{S,\mu}^{(st)} F_{S,\mu}^{(st)},\\
&F_{S,\mu}^{(st)} = \frac{1}{2}\,\sum_{i\not=j}\,\exp \left[ -a_{S,\mu}^{(st)} r_{ij}^2\right] P_{ij}^{(st)},
\end{split}
\label{eq:FS_gs}
\end{eqnarray}
where the coefficient $C_{\mu}^{(st)}$ are chosen as a real number in the present study.
Consequently the binding energy per particle in nuclear matter $B_{N=1}$ given in Eq.~(\ref{eq:1st_TOFS_cal}) in the 1st order TOFS calculation is presented as a quadratic equation with respect to the expansion coefficients $\left\{C_{S,\mu}^{(st)}\right\}$ in the Gaussian expansion of the correlation function $F_S$,
\begin{eqnarray}
\begin{split}
-B_{N=1} &= \frac{1}{A}\, \biggl[ {\langle \Phi_0 | \mathcal{H} | \Phi_0 \rangle}_{\rm c} + \sum_{s,t,\mu}  {\left\langle \Phi_0 \left| F_{S,\mu}^{(st)} \mathcal{H} + \mathcal{H} F_{S,\mu}^{(st)} \right| \Phi_0 \right\rangle}_{\rm c}\, C_{S,\mu}^{(st)} \\
& + \sum_{s,t,\mu} \sum_{s',t',\mu'} {\left\langle \Phi_0 \left|\, F_{S,\mu}^{(st)}\, \mathcal{H}\, F_{S,\mu'}^{(s't')} \right| \Phi_0 \right\rangle}_{\rm c}\, C_{S,\mu}^{(st)} C_{S,\mu'}^{(s't')} \biggr].  
\end{split}
\label{eq:B_N=1}
\end{eqnarray}
In the present study the size parameters $\left\{a_{S,\mu}^{(st)}\right\}$ are appropriately chosen under the consideration of the range of the nuclear force and Fermi wave number $k_F$, and thus the variational parameters are only $\left\{C_{S,\mu}^{(st)}\right\}$.
The variational condition in Eq.~(\ref{eq:system_eq}) gives simultaneous linear equations with respect to the set $\left\{C_{S,\mu}^{(st)}\right\}$~\cite{yamada18_cluster} as
\begin{eqnarray}
\begin{split}
& \sum_{s',t',\mu'} {\left\langle \Phi_{0} \left|  F_{S,\mu}^{(st)} \mathcal{H} F_{S,\mu'}^{(s't')}  \right| \Phi_{0} \right\rangle}_{\rm c}\, C_{S,\mu'}^{(s't')} 
= - \frac{1}{2}\, {\left\langle \Phi_{0} \left| F_{S,\mu}^{(st)} \mathcal{H} + \mathcal{H} F_{S,\mu}^{(st)} \right| \Phi_{0} \right\rangle}_{\rm c}.
\end{split}
\label{eq:linear_eq}
\end{eqnarray}
Solving these linear equations, we can obtain a unique solution for the set $\{C_{S,\mu}^{(st)}\}$.
As a result, we can evaluate the binding energy per particle in nuclear matter, $B_{N=1}$, and the radial behavior of the correlation functions, $f_S^{(st)}(r)$, in the 1st order TOFS method.

\begin{figure}[t]
\centering
\includegraphics*[width=0.6\hsize]{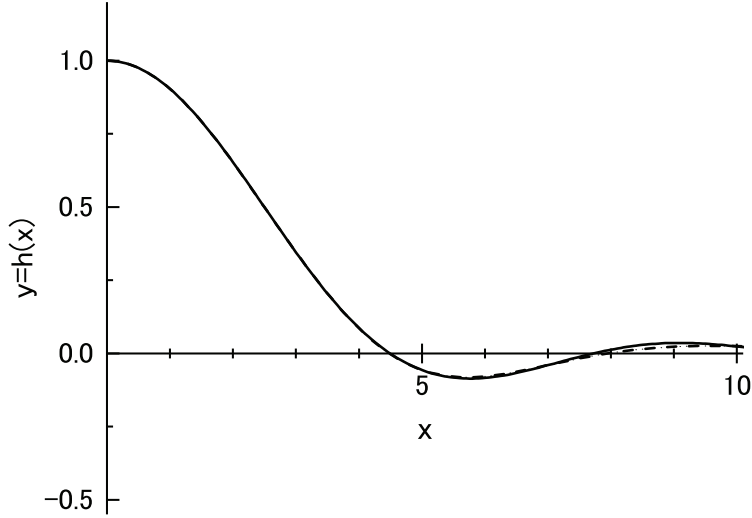}
\caption{
Compariosn with $h(x)=3j_{1}(x)/x$ (real line) and its Gaussian expansion in the right hand in Eq.~(\ref{eq:gs_h(x)}) (dashed line).
See the text.
}
\label{fig:h(x)_plot}
\end{figure} 

In Eq.~(\ref{eq:mt_123456}) we have to perform the multi-dimensional integral ($3\times 5=15$ dimensions).
Some integrals in Eq.~(\ref{eq:mt_123456}) can analytically reduce their dimensions by using the following formulae etc.,
\begin{eqnarray}
\begin{split}
&\int d\vc{r}\,h(k_F r) = \frac{4}{\rho}, \\ 
&\int d\vc{r}\,h(k_F r)\, h(k_F | \vc{r} - \vc{x}|) = \frac{4}{\rho}\,h(k_F x).
\end{split}
\label{eq:h(x)_formulae}
\end{eqnarray}
where $h(x)$ is defined in Eq.~(\ref{eq:h(x)}).
On the other hand, the Gaussian expansion of the function $h(x)$ may be useful to calculate the mutidimensional integral, 
\begin{eqnarray}
h(x) = \frac{3j_1(x)}{x} = \sum_{n} b_n \exp(-\nu_{n}x^2),
\label{eq:gs_h(x)}
\end{eqnarray}
when the formulae in Eq.~(\ref{eq:h(x)_formulae}) can not be applicable.
Figure~\ref{fig:h(x)_plot} shows the calculated result of the Gaussian expansion of $h(x)$ with use of Eq.~(\ref{eq:gs_h(x)}).
We see a relatively good reproduction of the damped oscillation of $h(x)$.   
This means that the Gaussian expansion of $h(x)$ works reasonably well, although one must be careful for the numerical precision.
It is known that the radial part of the central force $V_C$, $v_{C}^{(st)}(r)$, in Eq.~(\ref{eq:hamiltonian_central}) can be also expanded reasonably in terms of the Gaussian functions. 
Since the radial part of the correlation function $F_S$ is expanded in terms of the Gaussian functions, we can perform the nuclear matter calculation with the Gaussian integrals in the present TOFS framework.

\section{Results and discussion}
\label{sub:results_and_discussion}.
 
\begin{figure}[t]
\centering
\includegraphics*[width=0.8\hsize]{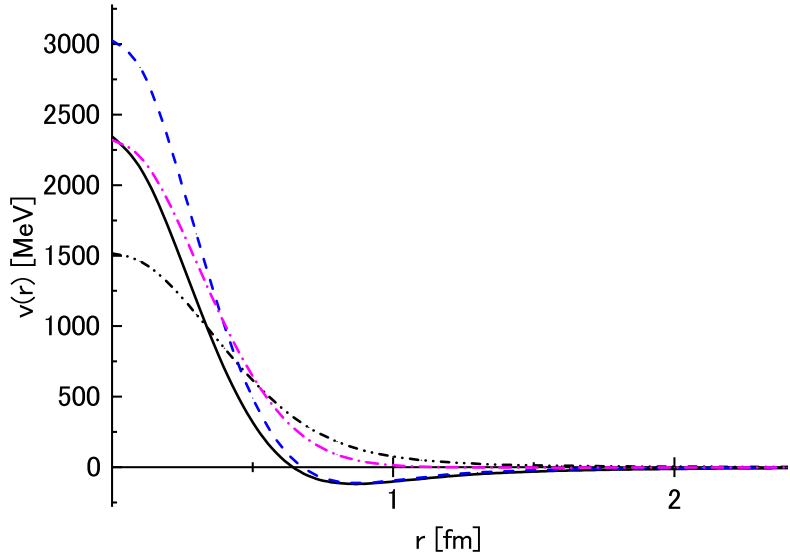}
\caption{
(Color)~Radial behaviors of the Argnne $V_{4}$' potential, $v_{C}^{(st)}(r)$:~$^{3}E$ (real line), $^{1}E$ (dash), $^{3}O$ (dash-dotted), and $^{1}O$ (dash-two-dotted).
}
\label{fig:av4_pot}
\end{figure}

We use the Argonne V4' potential for the $NN$ central force~\cite{wiringa95}.
This potential reproduces the binding energy of the deuteron in the $^{3}S_1$ channel, and the $NN$ phase shifts of the $^{1}S_{0}$, $^{3}S_{1}$, and $^{1}P_{1}$ channels are reasonably reproduced up to energies of about 350 MeV, while those of $^{3}S_{3,2}$ and $^{3}P_{0,1,2}$ are not well reproduced because of no tensor coupling etc.
The radial behavior of the Argonne $V_{4}$' potential, $v_C^{(st)}(r)$ in Eq.~(\ref{eq:hamiltonian_central}), are shown in Fig.~\ref{fig:av4_pot}.
The repulsive strength of the potential at $r=0$ is as strong as a couple of GeV for each spin-isospin channel.

In the present paper, the radial part of the spin-isospin dependent central correlation functions, $f^{(st)}_{S}(r)$, in Eqs.~(\ref{eq:exp_gs_FS}) or (\ref{eq:FS_gs}) are expanded into four Gaussian functions with the size parameters, $a^{(st)}_{S,\mu}=1.2, 1.8, 2.4$, and $3.0$ fm$^{-2}$, for each spin-isospin channel and each density. 
Then, the simultaneous linear equations with respect to sixteen variables concerning the expansion coefficients $C^{(st)}_{S,\mu}$ in Eq.~(\ref{eq:linear_eq}) are solved.
As a result we can evaluate the energy per particle for nuclear matter in Eq.~(\ref{eq:B_N=1}) and the radial behavior of each spin-isospin correlation functions.  

The density dependence of the energy per particle for symmetric nuclear matter, $E/A=-B_{N=1}$, with the 1st order TOFS calculation is presented in Table~\ref{tab:1}.
It is found that the results of the present calculation are reasonably reproduced, compared with those of the BHF approach~\cite{baldo12}, although a couple of MeV attraction is deficient in lower density ($\rho=0.03$ and $0.05$ fm$^{-3}$) and a few MeV overbinding is seen at higher density ($\rho=0.20$ fm$^{-3}$).
As mentioned in Sec.~\ref{sec:Introduction}, the density dependence of $E/A$ for symmetric nuclear matter on the calculated methods such as BBG, SCGF, AFDMC, and FHNC is almost the same as that of BHF in the case of the AV4' potential, but the difference among them is slightly enhanced (a few MeV) in the region of $\rho \ge \rho_0=0.17$~fm$^{-3}$ (see Fig.~3 in Ref.~\cite{baldo12}). 
Therefore, the overbinding by a few MeV at $\rho=0.20$ fm$^3$ is likely to be within the difference among the calculated methods.
However, the shortage of the attraction at the lower density may be due to the fact that the present calculation is the lowest order one in the TOFS framework. 
The 2nd order TOFS calculation using Eq.~(\ref{eq:2nd_TOFS_cal}) is considered to be able to recover the small shortage.

\begin{table}[t]
\caption{Density dependence of the energy per particle for symmetric nuclear matter obtained by the 1st order TOFS method with the Argonne $V_4$' potential, compared with those of the BHF approach~\cite{baldo12}. 
The energy is given in the unit of MeV.}
\begin{center}
\begin{tabular}{cccccc}
\hline
$\rho$ [fm${^{-3}}$]  & \ 0.03 & \ 0.05 & \ 0.10 & \ 0.17 & \ 0.20 \\
\hline
1st order TOFS  & $-5.2$ & $-8.5 $ & $-16.1$ & $-26.8$  & $-32.9$ \\
BHF~\cite{baldo12} &  $-7.4$   &  $-11.4$ & $-17.7$ & $-26.4$ & $-29.7$ \\
\hline
\end{tabular}
\end{center}
\label{tab:1}
\end{table}


\begin{table}[t]
\caption{Contributions from various matrix elements in Eq.~(\ref{eq:1st_TOFS_cal}) for the energy per particle for nuclear matter at each density. 
The energy is given in the unit of MeV.
See text.}
\begin{center}
\begin{tabular}{ccccccccccc}
\hline
$\rho$ [fm${^{-3}}$] 
& ~~
& $\frac{1}{A}{\langle T \rangle}_{\rm c}$ & $\frac{1}{A}{\langle V_C \rangle}_{\rm c}$ & $\frac{1}{A}{\langle F\mathcal{H}+\mathcal{H}F \rangle}_{\rm c}$ & $\frac{1}{A}{\langle F\mathcal{H}F \rangle}_{\rm c}$ 
& ~~
& $E^{(0)}$
& $E^{(2b)}$
& $E^{(mb)}$
& $E$
\\
\hline
0.100 & & 16.2 & $-16.4$ & $-31.6$ & 15.8  & & ~$-0.3$~ & ~$-10.0$~ & ~$-5.8$~ & $-16.1$ \\ 
0.170 & & 23.0 & $-21.9$ & $-55.8$ & 27.9  & & ~$\ \ 1.1$~ & ~$-21.0$~ & ~$-6.9$~ & $-26.8$ \\
0.200 & & 25.7 & $-23.6$ & $-70.1$ & 35.0  & & ~$\ \ 2.1$~ & ~$-26.9$~ & ~$-8.1$~ & $-32.9$ \\ 
\hline
\end{tabular}
\end{center}
\label{tab:2}
\end{table}

The contributions from the matrix element at each density, which are given in Eq.~(\ref{eq:1st_TOFS_cal}), are shown in Table~\ref{tab:2}, where we define $\frac{1}{A}{\langle \mathcal{O} \rangle}_{\rm c} \equiv \frac{1}{A}{\langle \Phi_0 | \mathcal{O} | \Phi_0 \rangle}_{\rm c}$ with $\mathcal{O}=T, V_{C}, F\mathcal{H}+\mathcal{H}F$, and $F\mathcal{H}F$.
In the uncorrelated nuclear matter wave function $\Phi_0$ in Eq.~(\ref{eq:femi_gas_wf}), the energy per particle for nuclear matter, $E/A=\frac{1}{A}{\langle T + V_C \rangle}_{\rm c}$, is almost zero or slightly positive at each density.
However, the correlations in each nucleon pair have a great effect to bound nucleons in nuclear matter. 
In fact, the sum of $\frac{1}{A}{\langle F\mathcal{H}+\mathcal{H}F \rangle}_{\rm c}$ and $\frac{1}{A} {\langle F\mathcal{H}F \rangle}_{\rm c}$ gives almost all parts of the energy per particle for nuclear matter, as shown in Table~\ref{tab:2}.

In the present framework, the following relation is exactly established at the stationary point:~$\frac{1}{A}{\langle F\mathcal{H}+\mathcal{H}F \rangle}_{\rm c} = -2 \times \frac{1}{A}{\langle F\mathcal{H}F \rangle}_{\rm c}$.
This is because in the 1st order TOFS framework the energy per particle for nuclear matter is given as the quadratic form with respect to the expansion coefficients $C^{(st)}_{S,\mu}$ (see Eq.~(\ref{eq:B_N=1})) and the stationary point is given by solving the simultaneous linear equations in Eq.~(\ref{eq:linear_eq}).
This relation is numerically realized in Table~\ref{tab:2}.
We expect that this relation is approximately established even in the 2nd order TOFS calculation using Eq.~(\ref{eq:2nd_TOFS_cal}), since the 1st order TOFS calculation gives the reasonable results in comparison with those of BHF, BBG, SCGF, AFDMC, and FHNC.

It is also interesting to see the contributions from the many body terms in Eq.~(\ref{eq:1st_TOFS_cal}), where $-B_{N=1}=E=\frac{1}{A}{\langle \Phi_0 | (1+F) \mathcal{H} (1+F) | \Phi_0 \rangle}_{\rm c}$. 
Here we define the following quantities:~$E^{(0)}=\frac{1}{A}\langle \mathcal{H} \rangle_{\rm c}$, $E^{(2b)}={\left[\frac{1}{A}\langle \Phi_0 | F\mathcal{H} + \mathcal{H}F + F\mathcal{H}F | \Phi_0 \rangle_{\rm c}\right]}^{(2b)}$, and $E^{(mb)}={\left[\frac{1}{A}\langle \Phi_0 | F\mathcal{H} + \mathcal{H}F + F\mathcal{H}F | \Phi_0 \rangle_{\rm c}\right]}^{(mb)}$ with $E=E^{(0)}+E^{(2b)}+E^{(mb)}$, where $E^{(2b)}$ denotes the contribution from only the two body terms in the matrix element of $\frac{1}{A}\langle \Phi_0 | F\mathcal{H} + \mathcal{H}F + F\mathcal{H}F | \Phi_0 \rangle_{\rm c}$, and  $E^{(mb)}$ stands for the sum of the contributions from the three body terms to six body ones in the matrix element.
The results are shown in Table~\ref{tab:2}. 
We notice that the contribution of $E^{(mb)}$ amounts to be about $30\ \%$ of the  energy per particle for nuclear matter, $-B_{N=1}=E$, at each density.
This fact means that the many body effects cannot be neglected in nuclear matter.
This importance of the many body terms has been emphasized in nuclear matter calculations with the FHNC method~\cite{pandharipande79,akmal98} and also pointed out in the ab initio calculations of finite nuclei with use of TOAMD~\cite{myo15,myo17_1,myo17_2,myo17_3,myo17_4,myo17_5,myo17_6,lyu17}.

\begin{figure}[t]
\centering
\includegraphics*[width=0.7\hsize]{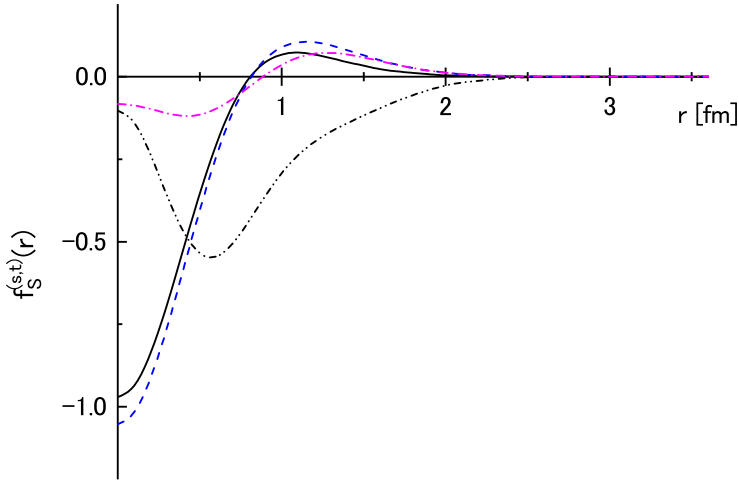}
\caption{
(Color)~Correlation functions $f^{(st)}_{S}(r)$ for the $^3E$ (real line), $^{1}E$ (dash), $^{3}O$ (dash-dotted), and $^{1}O$ (dash-two-dotted) channels at $\rho=0.17$ fm$^{-3}$.
}
\label{fig:cf_rho170}
\end{figure} 

The radial behaviors of the spin-isospin correlation functions, $f^{(st)}_S(r)$, at $\rho=0.17$ fm$^{-3}$ are shown in Fig.~\ref{fig:cf_rho170}.
For the $^{3,1}E$ channels, the correlation functions have negative values at short distance up to $r \simeq 0.8$~fm.
This behavior indicates a reduction of the short-range amplitude of the nucleon pair from the uncorrelated Fermi-gas wave function. 
Beyond $r \simeq 0.8$~fm, they become positive and give the maximum values at $r \simeq 1.1$~fm, having a long tail up to $r\simeq 2$~fm, corresponding to the intermediate and long distances of the nuclear force.
Thus, $f^{(st)}_{S}$ with the $^{3,1}E$ channels have two aspects:~One is the short-range correlation to reduce the repulsion in the nuclear force, and the other is the intermediate and long-range correlations to get the attraction from the nuclear force.

On the other hand, the contribution from the $^{3,1}O$ channels in the energy per particle for nuclear matter is in general smaller than that from the $^{3,1}E$ channels, because the relative orbital angular momentum $\ell$ of nucleon pair is $\ell \ge 1$ in the former channels, and the present AV4' potential has no tensor and spin-orbit components.
Consequently, the magnitudes of the correlation functions for the $^{3,1}O$ channels are smaller than those for the $^{3,1}E$ ones, as shown in Fig.~\ref{fig:cf_rho170}.
However, the $^{1}O$ channel correlation function has an interesting behavior: Its magnitude beyond $r \simeq 0.5$~fm is largest among other correlation functions. 
The reasons are given as follows:~The $^1O$ channel of the AV4' potential is repulsive in the whole region, and the magnitude of the repulsion is the smallest at $r=0$ among the four spin-isospin channels, but it is the largest beyond $r\simeq 0.5$~fm  (see Fig.~\ref{fig:av4_pot}). 
Then the $^1O$ channel correlation function has smaller negative values up to $r\simeq 0.5$~fm than those with the $^{3,1}E$ channels, while it has larger negative values beyond $r\simeq 0.5$~fm.
This behavior reduces the repulsive effect of the $^1O$ channel force beyond $r\simeq 0.5$~fm as well as the short-range region ($r < 0.5$~fm) in nuclear matter. 


\section{Summary}
\label{sub:summary}.

The tensor optimized Fermi sphere (TOFS) method was applied first for the study of the property of nuclear matter using the AV4' $NN$ potential. 
In the TOFS method, the correlated nuclear matter wave function is taken to be a power-series-type, $\Psi_N = [ \sum_{n=0}^{N} (1/n!)F^n ] \Phi_0$, called as the $N$th order TOFS wave function, where $\Phi_0$ and $F$, respectively, stand for the uncorrelated Fermi gas wave function and the correlation function that can induce central, spin-isospin, tenor, etc.~correlations. 
This expression has been ensured by a cluster expansion theory~\cite{yamada18_cluster}, and the energy per particle for nuclear matter ($E/A$) with use of $\Psi_N$ is presented as the sum of certain linked diagrams.
The radial parts of the spin-isospin dependent $F$ are expanded in terms of the Gaussian functions, and their expansion coefficients are optimally determined in the variation in $E/A$.  

In the present paper, the 1st order TOFS wave function $\Psi_{N=1}$ was used to evaluating the density dependence of $E/A$, and we took into account the contributions from all the many body terms (from one to six body term)  originating from the product of the nuclear matter Hamiltonian $\mathcal{H}$ and $F$.
In this TOFS method, the expansion coefficients of the Gaussian expansion for the correlation functions are determined by solving the simultaneous linear equations with respect to the expansion coefficients.  
We found that the density dependence of $E/A$ is reasonably reproduced, in comparison with other methods such as the BHF approach etc., and the spin-isospin dependent central  correlation functions are described nicely by the Gaussian expansion method.

The present results indicate that the TOFS method is useful to study the nuclear matter with the bare interaction among nucleons.
As mentioned in Sec.~\ref{sec:Introduction}, the significant dependence of the density dependent behavior of $E/A$ on the calculated methods (BHF, BBG, SCGF, FHNC, AFDMC) has been reported in the cases of the AV6' and AV8' potentials, even in the lower density region including $\rho=\rho_0$.
Therefore, it is an intriguing subject to study the density dependence of $E/A$ with use of the AV6' and AV8' potentials with the 1st and 2nd order TOFS methods etc.
We will report their results elsewhere in near future.

\section*{Acknowledgments}

This work was partially supported  by the JSPS KAKENHI Grans No.~26400283 and No.~18K03629.



\begin{thebibliography}{99}

\bibitem{glendenning}
N.~K.~Glendenning, Compact Stars:\ Nuclear Physics, Particle Physics, and General Relativity (Springer, 1996).
\bibitem{raffelt96}
G.~G.~Raffelt, Stars as Laboratories for Fundamental Physics:\ The Astrophysics of Neutrinos, Axions, and Other Weakly Interacting Particles (University of Chicago, Chicago, 1996).

\bibitem{roepke98}
G.~R\"opke, A.~Schnell, P.~Schuck, P.~Nozi\'eres, Phys. Rev. Lett. {\bf 80}, 3177 (1998).
\bibitem{beyer00}
M.~Beyer, S.~A.~Sofianos, C.~Kuhrts, G.~R\"opke, and P.~Schuck, Phys. Lett. B {\bf 488}, 247 (2000).
\bibitem{takemoto04}
H.~Takemoto, M.~Fukushima, S.~Chiba, H.~Horiuchi, Y.~Akaishi, A.~Tohsaki, Phys.Rev. C {\bf 69}, 035802 (2004).
\bibitem{sogo09}
T.~Sogo, R.~Lazauskas, G.~R\"opke, P.~Schuck, Phys. Rev. C {\bf 79}, 051301(R) (2009).
\bibitem{sogo10-1}
T.~Sogo, G.~R\"opke, P.~Schuck, Phys. Rev. C {\bf 81}, 064310 (2010)
\bibitem{sogo10-2}
T.~Sogo, G.~R\"opke, P.~Schuck, Phys. Rev. C {\bf 82}, 034322 (2010).

\bibitem{wildermuth77}
K.~Wildermuth and Y.~C.~Tang, A Unified Theory of the Nucleus (Vieweg, Braunschweig, 1977).
\bibitem{ikeda80}
K.~Ikeda, H.~Horiuchi, and S.~Saito, Prog. Theor. Phys. Supplement {\bf 68}, 1 (1980).
\bibitem{tohsaki00}
A.~Tohsaki, H.~Horiuchi, P.~Schuck, and G.~R\"opke (2001), Phys. Rev. Lett. {\bf 87}, 192501.
\bibitem{yamada11}
T.~Yamada, Y.~Funaki, H.~Horiuchi, G.~R\"opke, P.~Schuck, and A.~Tohsaki, in Cluster in Nuclei - Vol. 2 edited by C. Beck (Springer-Verlag, Berlin, 2011), 848, Chap. 5.
\bibitem{horiuchi12}
H.~Horiuchi, K.~Ikeda, and K. Kato,  Prog. Theor. Phys. Supplement {\bf 192}, 1 (2012).

\bibitem{brueckner55_1}
K.~A.~Brueckner and C.~A.~Levinson, Phys. Rev. {\bf 97}, 1344 (1955).
\bibitem{brueckner55_2}
K.~A.~Brueckner, Phys. Rev. {\bf 100}, 36 (1955).
\bibitem{brueckner58}
K.~A.~Brueckner and J.~L.~Gammel, Phys. Rev. {\bf 109},1023 (1958).

\bibitem{bethe56}
H.~A.~Bethe, Phys. Rev. {\bf 103}, 1353 (1956).
\bibitem{bethe57}
H.~A.~Bethe and J.~Goldstone, Proc. Roy. Soc. {\bf A238}, 551 (1957).
\bibitem{goldstone57}
J.~Goldstone, Proc. Roy. Soc. {\bf A239}, 267 (1957).

\bibitem{mahayux}
C.~Mahaux and R.~Sartor, in Nuclear Matter and Heavy Ion Collisions, edited by M. Soyeur et al., NATO Advanced
Study Institute Ser. B, Vol. 205 (Plenum Press,New York, 1989).
\bibitem{baldo91}
M.~Baldo, I.~Bombaci, L.~S.~Ferreira, G.~Giansiracusa, and U.~Lombardo, Phys. Rev. C {\bf 43}, 2605 (1991).
\bibitem{schulze95}
H.-J.~Schulze, J.~Cugnon, A.~Lejuene, M.~Baldo, and U.~Lombardo, Phys. Rev. C {\bf 52}, 2785 (1995).
\bibitem{bradow96}
B.~H.~Brandow, Phys. Rev. 152, 863 (1996).

\bibitem{day67}
B.~D.~Day, Rev. Mod. Phys. {\bf 39}, 719 (1967), and references therein.
\bibitem{day83}
B.~D.~Day, in Brueckner-Bethe Calculations of Nuclear Matter, Proceedings of the International School of Physics
"Enrico Fermi" Course LXXIX, edited by A.~Molinari (Editrice Compositori, Bologna, 1983), pp.~1-72.

\bibitem{song98}
H.~Q.~Song, M.~Baldo, G.~Giansiracusa, and U.~Lombardo, Phys. Rev. Lett. {\bf 81}, 1584 (1998).
\bibitem{baldo00}
M.~Baldo, G.~Giansiracusa, U.~Lombardo, and H.~Q.~Song, Phys. Lett. B {\bf 473}, 1 (2000);
\bibitem{baldo01}
M.~Baldo, A.~Fiasconaro, H.~Q.~Song, G.~Giansiracusa, and U.~Lombardo, Phys. Rev. C {\bf 65}, 017303 (2001)
\bibitem{sartor06}
R.~Sartor, Phys. Rev. C {\bf 73}, 034307 (2006).

\bibitem{dickhoff04}
W.~H.~Dickhoff and C. Barbieri, Prog. Part. Nucl. Phys. {\bf 52}, 377 (2004).
\bibitem{frick05}
T.~Frick, H.~M\"uther, A.~Rios, A.~Polls, and A.~Ramos, Phys. Rev. C {\bf 71}, 014313 (2005).
\bibitem{soma06}
V. Som\`a and P.~Boz\.ek, Phys. Rev. C {\bf 74}, 045809 (2006); {\bf 78}, 054003 (2008).
\bibitem{rios09}
A.~Rios, A.~Polls, and I.~Vida\~na, Phys. Rev. C {\bf 79}, 025802 (2009).

\bibitem{gandolfi07}
S.~Gandolfi, F.~Pederiva, S.~Fantoni, and K.~E.~Schmidt, Phys. Rev. Lett. {\bf 98}, 102503 (2007).
\bibitem{gandolfi09}
S.~Gandolfi, A.~Y.~Illarionov, K.~E.~Schmidt, F.~Pederiva, and S.~Fantoni, Phys. Rev. C {\bf 79}, 054005 (2009).

\bibitem{carlson03}
J.~Carlson, J.~Morales, V.~R.~Pandharipande, and D.~G.~Ravenhall, Phys. Rev. C {\bf 68}, 025802 (2003).

\bibitem{iwamoto57}
F.~Iwamoto and M.~Yamada, Prog. Theor. Phys. {\bf 17}, 543 (1957).
\bibitem{fantoni72}
S.~Fantoni and S.~Rosati, Nuovo Cimento A {\bf 10}, 145 (1972).
\bibitem{fantoni74}
S.~Fantoni and S.~Rosati, Nuovo Cimento A {\bf 20}, 179 (1974).
\bibitem{fantoni78}
S.~Fantoni and S.~Rosati, Nuovo Cimento A {\bf 43}, 413 (1978).
\bibitem{pandharipande79}
V.~R.~Pandharipande and R. B. Wiringa, Rev. Mod. Phys. {\bf 51}, 821 (1979).
\bibitem{akmal98}
A.~Akmal, V.~R.~Pandharipande, and D.~G.~Ravenhall, Phys. Rev. C {\bf 58}, 1804 (1998).

\bibitem{baardsen13}
G. Baardsen, A. Ekstr\"om, G. Hagen, and M. Hjorth-Jensen, Phys. Rev. C {\bf 88}, 054312 (2013).
\bibitem{hagen14}
G.~Hagen, T.~Papenbrock, M.~Hjorth-Jensen, and D.~J.~Dean, Rep. Prog. Phys. {\bf 77}, 096302 (2014).

\bibitem{abe09}
T.~Abe and R.~Seki, Phys. Rev. C {\bf 79}, 054002 (2009).

\bibitem{wiringa95}
R.~B.~Wiringa, V.~G.~J.~Stokes, R.~Schiavilla, Phys. Rev. C {\bf 51}, 38 (1995).

\bibitem{pudliner95}
B.~S.~Pudliner, V.~R.~Pandharipande, J.~Carlson, R.~B.~Wiringa, Phys. Rev. Lett. {\bf 74}, 4396 (1995).

\bibitem{brockmann90}
R.~Brockmann and R.~Machleidt, Phys. Rev. C {\bf 42}, 1965 (1990).

\bibitem{baldo12}
M.~Baldo, A.~Polls, A.~Rios, H.-J.~Schulze, and I.~Vida\~na, Phys. Rev. C {\bf 86}, 064001 (2012). 

\bibitem{myo15}
T.~Myo, H.~Toki, K.~Ikeda, H.~Horiuchi, and T.~Suhara, Prog. Theor. Exp. Phys. {\bf 2015}, 073D02 (2015).
\bibitem{myo17_1}
T.~Myo, H.~Toki, K.~Ikeda, H.~Horiuchi, and T.~Suhara, Phys. Lett. B, {\bf 769}, 213 (2017).
\bibitem{myo17_2}
T.~Myo, H.~Toki, K.~Ikeda, H.~Horiuchi, and T.~Suhara, Phys. Rev. C {\bf 95}, 044314 (2017).
\bibitem{myo17_3}
T.~Myo, H.~Toki, K.~Ikeda, H.~Horiuchi, and T.~Suhara, Prog. Theor. Exp. Phys. {\bf 2017}, 073D01 (2017).
\bibitem{myo17_4}
T.~Myo, H.~Toki, K.~Ikeda, H.~Horiuchi, and T.~Suhara, Phys. Rev. C {\bf 96}, 034309 (2017).
\bibitem{myo17_5}
T.~Myo, H.~Toki, K.~Ikeda, H.~Horiuchi, T.~Suhara, M.~Lyu, M.~Isaka, and T.~Yamada, Prog. Theor. Exp. Phys. {\bf 2017}, 111D01 (2017).

\bibitem{enyo03}
Y.~Kanada-En'yo, M.~Kimura, and H.~Horiuchi, C. R. Phys. {\bf 4}, 497 (2003).
\bibitem{enyo12}
Y.~Kanada-En'yo, M.~Kimura, and A.~Ono, Prog. Theor. Exp. Phys. {\bf 2012}, 01A202 (2012).

\bibitem{myo17_6}
T.~Myo, Prog. Theor. Exp. Phys., {\bf 2018}, 031D01 (2018),
\bibitem{lyu17}
M.~Lyu, M.~Isaka, T.~Myo, H.~Toki, K.~Ikeda, H.~Horiuchi, T.~Suhara, and T.~Yamada, Prog. Theor. Exp. Phys. {\bf 2018}, 011D01 (2018).

\bibitem{kummel78}
H.~K\"ummel, H.~K.~L\"uhrmann, and J.~G.~Zabolitzky, Phys. Rep. {\bf 36} 1, (1978).
\bibitem{barlett81}
R.~Barlett, Ann. Rev. Phys. Chem. {\bf 32}, 359 (1981).
 

\bibitem{yamada18_cluster}
T.~Yamada, ``Tensor optimized Fermi sphere method for nuclear matter -- power series correlated wave function and a cluster expansion  --'', arXiv:1808.07257.

\bibitem{co94}
G. C\'{o}, A. Fabrocini, and S. Fantoni, Nucl. Phys. A {\bf 568}, 73 (1994).

\bibitem{arias07}
F.~Arias de Saavedra, C. Bisconti, G. C\'{o}, A. Fabrocinic, Phys. Rep. {\bf 450},1  (2007).

\end{thebibliography}
%




\end{document}